\documentclass[a4paper,pre,12pt,lcy,amsmath,amssymb,graphicx,nofootinbib,showpacs,superscriptaddress,floatfix]{revtex4}
\usepackage{amsmath, mathrsfs, latexsym, amssymb, color, graphicx, comment, hyperref, scrhack}
\usepackage{lmodern}
\usepackage{physics}
\usepackage{amssymb}

\newcommand{\dfr}[2]{\frac {\displaystyle #1}{\displaystyle #2}}
\begin{document}

\title{To understanding of slow and non-monotonic relaxation in Al--Y eutectic melts}

\author{M.\,G.\,Vasin}
\address{Institute for High Pressure Physics of Russian Academy of Sciences, 142190 Moscow, Russia}
\address{Ural Federal University, Ekaterinburg, Russia}
\author{V.\,G.\,Lebedev}
\address{Research center of metallurgical physics and materials science of UdmFRC UrB RAS, 426067 Izhevsk, Russia}
\address{Department of Mathematics, Informatics and Physics, Udmurt State University, 426034 Izhevsk, Russia}

\begin{abstract}
We discuss the nature of the slow relaxation processes in glass-forming eutectic melts right after melting. For specific, we focus on the binary metallic melt Al--Y, which in addition to the slow relaxation shows unusual non-monotonic dynamics. We argue this slow dynamics is an result of non-linearity of diffusion processes in initially non-homogenous sample, and the nature of slow relaxation processes in eutectic melts after melting is similar to the nature of spinodal decomposition, when reason for the slowdown is the thermodynamic instability. To support this assertion we considered the model with combined Gibbs potential of the Al-Y liquid solution, in which the presence of the stoichiometric phase remains is taken into account. We show that in this system the instability mathematically described by the Cahn--Hilliard type equation can develop, and that fluctuation accounting in the considered model allows qualitatively describe the non-monotonic relaxation observed in the Al-based nonequilibrium melts.
\end{abstract}

\maketitle

%\begin{pacs} \pacs{PACS numbers: 05.70.Fh; 05.70.Ln; 02.30.Jr; 02.70.-c } \end{pacs}

%%%%%%%%%%%%%%%%%%%%%%%%%%%%%%%%%%%%%%%%%%%%%%%%%%%%%%%%%%%%%%%%%%%%%%%
\section{Introduction}
\label{sec:intro}

\begin{figure}[h!]
\centering
\includegraphics[width=1\textwidth]{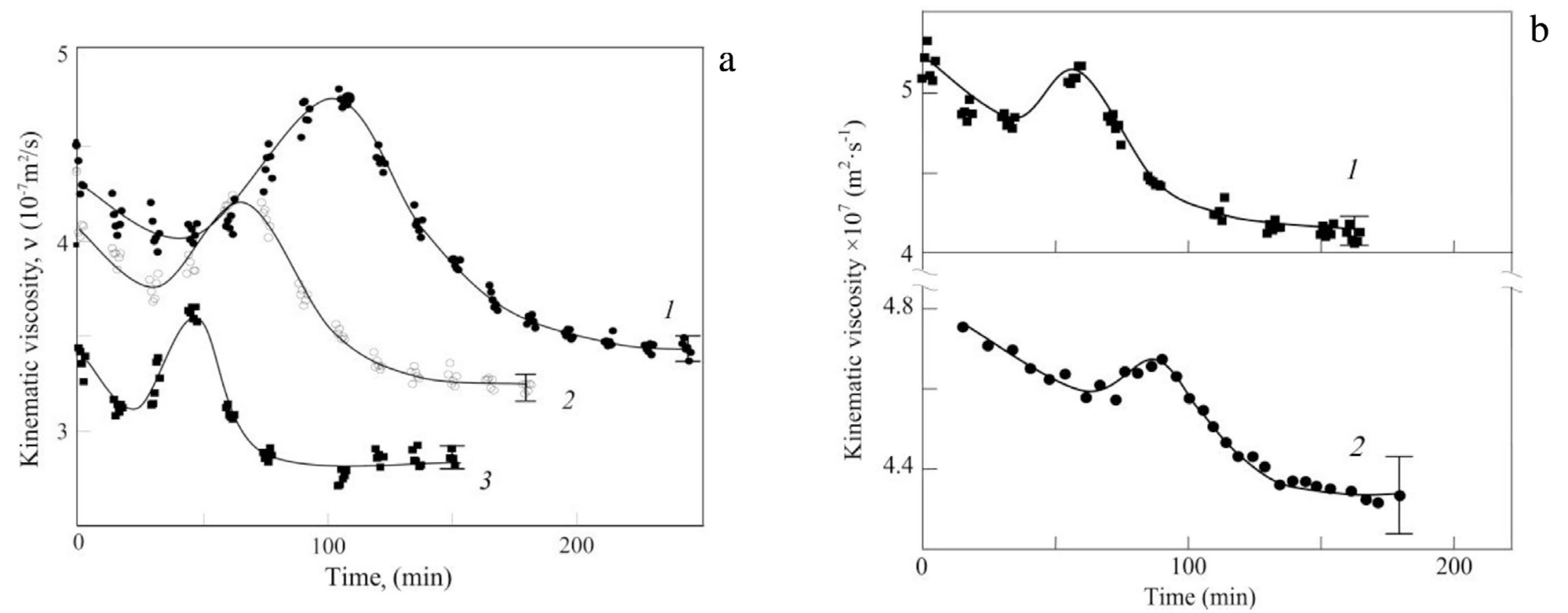}
\caption{The time dependencies of the Al-based melts viscosity: a) Al$_{87}$Ni$_{8}$Y$_5$, at 900$^{\circ}$C (1), 1050$^{\circ}$C (2) and 1200$^{\circ}$C (3), obtained after heating from room temperature; b) The time dependencies of the liquid melt viscosity, Al$_{86}$Ni$_8$La$_6$ (1) and Al$_{86}$Ni$_{8}$Ce$_6$ (2), at $1100^{\circ}$C \cite{4}}
\label{fig1}
\end{figure}

Some well known physical phenomena, observed in metallurgical processes, raise questions to physicists so far. The reason of these questions is the absence of a reliable description of these phenomena in terms of a generally accepted theory. One of these phenomena is the slow relaxation of some glass-forming metal melts after melting~\cite{1,2}, which relaxation time reaches few hours. In some cases this relaxation is accompanied by an unusual non-monotonic dependence of the melt viscosity \cite{2, 3, 4, 5} on time (see Fig.\,\ref{fig1}). In metallurgy these effects are called as remelting and explained as the result of slow dissolution of refractory solid phase fragments in liquid. However, the kinetics of these relaxation processes cannot be explained in terms of the linear diffusion model, since simple estimations give the relaxation time values order of some seconds.

Usually the slow relaxation is observed in structure-sensitive properties, like viscosity or resistivity. Large-scale investigation of this phenomenon in Fe-based and Al-based melts was undertaken at the end of the 20-th century. As a result, significant progress in understanding the physics of this process was achieved. In particular, it was concluded, that the slowdown is accompanied by the prolonged inhomogeneity retention in form of metastable microemulsion formation, in which droplets of 10--100 angstroms in size exist for a long time~\cite{PB}. This conclusion was also supported by the direct structure investigations \cite{S1,S2}.

At present time the investigations are continued. In particularly not long ago in \cite{Gasser} it was reported that Bi-In has a tendency to homocoordination. It was shown that to obtain a macroscopic homogeneous liquid alloy characterized by a stable resistivity it was necessary to wait several days and to heat to 700 C$^{\circ}$ above the melting point of the elements. Great interest today is attracted by Al-based eutectic alloys. As it is note above, in the melts of these systems, specifically in Al--Y, Al--La and Al--Ce melts (see Fig.\,\ref{fig1}), one can observe not only slow relaxation but also non-monotonic relaxation processes.  When, for a certain period of time after melting, the melt viscosity decreases exponentially, but at the some point it suddenly begins to grow, reaching local maximum, and then returns to the normal exponentially decreasing mode.

In this paper we will try to demolish seeming mysticism of these phenomena, founding on well known theoretical ideas and conceptions about relaxation processes kinetics in melts, based on the Cahn--Hilliard expressions. Also we will discuss alternative approach to description of these phenomena, based on modified kinetic Ising model.

As an example we consider the post melting relaxation of Al--Y-based melts. The initial inhomogeneities Al$_3$Y and Al$_2$Y  have characteristic size order $10^{-5}$ m. Estimation of characteristic dissolution time these inhomogeneities was done in the paper \cite{VMI} and is $10^{-2}$ s, which is many fewer than the relaxation time, observed on the experiment: $\tau\approx ~10^{4}$ s. It was experimentally established that in the melts of Al--Y alloys the non-monotonic viscosity behavior is observed both in the presence of other solutes and without them. One can assume that the nature of this unusual phenomenon is mainly related to the peculiarities of Al--Y melt relaxation. Therefore, initially for its description, we can limit ourselves to the consideration of the binary melt.

%%%%%%%%%%%%%%%%%%%%%%%%%%%%%%%%%%%%%%%%%%%%%%

%%%%%%%%%%%%%%%%%%%%%%%%%%%%%%%%%%%%%%%%%%%%%%%
%\begin{figure}[htb]
%\begin{center}
%\includegraphics[width=0.6\textwidth]{fig2.pdf}
%\caption{The temporal dependencies of the viscosity of Al$_{95}$Y$_5$ (a) and Al$_{90}$Y$_{10}$ (b) melts at the various temperatures \cite{4,5}}
%\end{center}
%\label{fig2}
%\end{figure}
%%%%%%%%%%%%%%%%%%%%%%%%%%%%%%%%%%%%%%%%%%%%%%%

\begin{figure}[h!]
\centering
\includegraphics[width=0.8\textwidth]{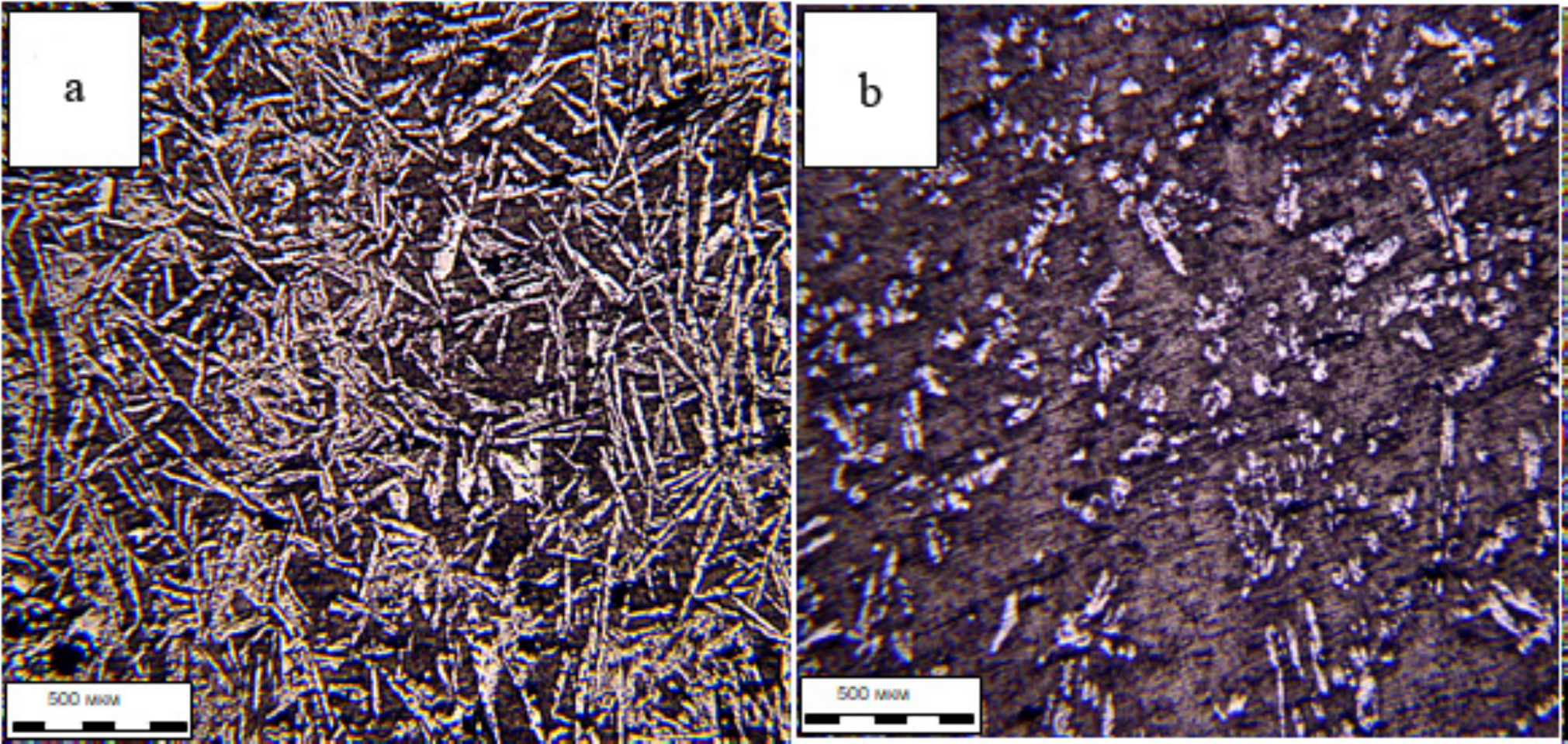}
\caption{Ligatures of initial samples a) Al--Y10\%, b) Al--Y5\% (from \cite{RMET}). In the structure one can see the inclusions of stoichiometric Al$_3$Y compounds which characteristic size $\sim 10^{-5}$ m.}
\label{LIG}
\end{figure}

As a possible explanation of the abnormally slow relaxation and non-monotonic dependence of viscosity on time in \cite{VMI} it was assumed that they are related with nonlinearity of the concentration dependence of the system chemical potential near the liquidus line.
Indeed, the melts right after melting are non-homogenous because of  non-homogeneity of initial solid alloy (see Fig.\,\ref{LIG}).  One can suppose that in ``young''  melt the yttrium atoms do not tend to leave at once the regions with high its concentration which stay ``energetically attractive'' long time. The effective Gibbs potential of this system is combination of the potentials of liquid Al--Y solution and stoichiometric Al$_2$Y and Al$_3$Y compounds. We suppose this can leads to strong nonlinearity of relaxation kinetics of non-homogeneous melt right after it melting.

%%%%%%%%%%%%%%%%%%%%%%%%%%%%%%%%%%%%%%%%%%%%%%
\begin{figure}[htb]
\centering
\includegraphics[width=0.6\textwidth]{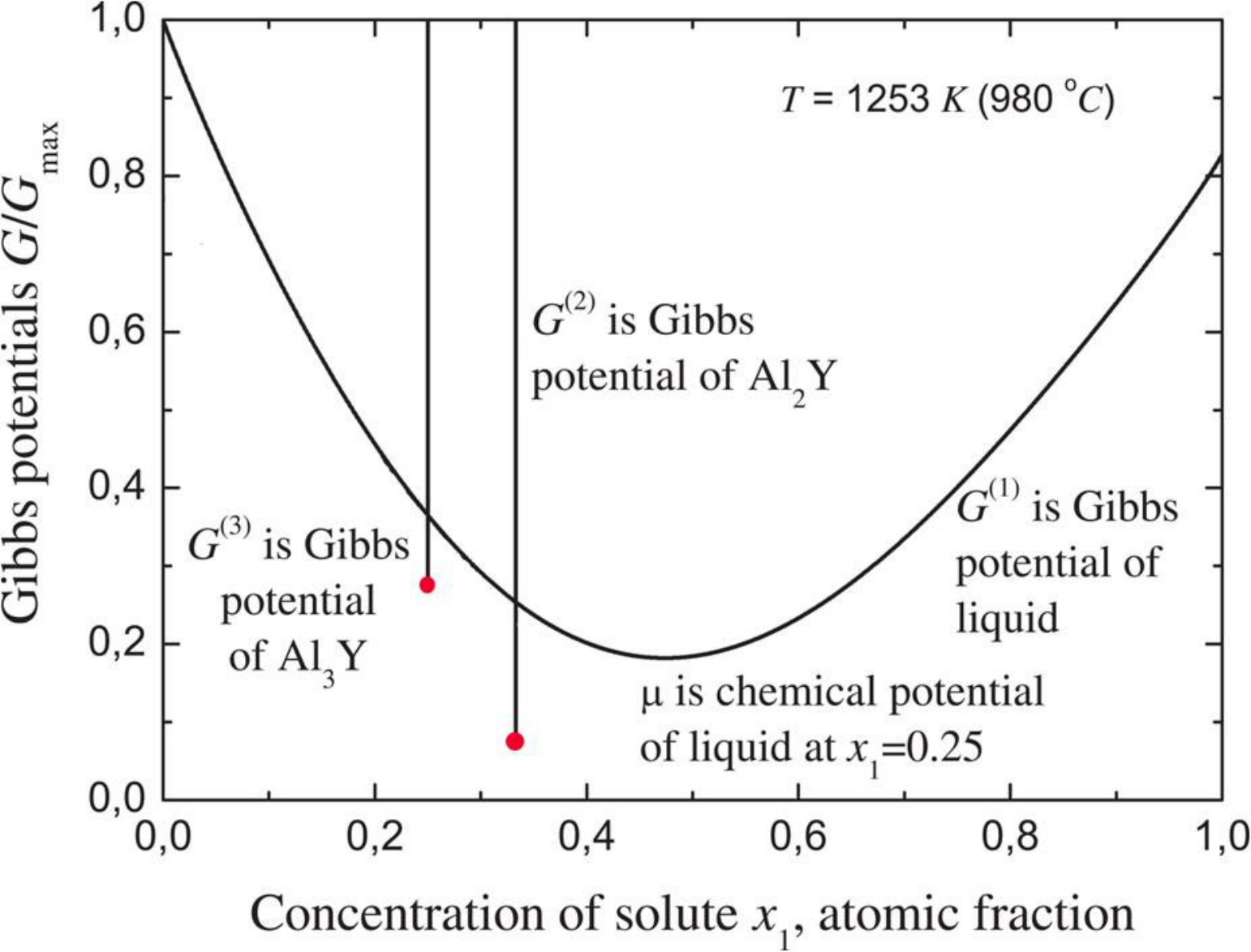}
\caption{Gibbs potentials of the liquid phase and two stoichiometric phases, Al$_3$Y and Al$_2$Y, in Al--Y alloy at the temperature $T=980^{\circ}$C from the Computational Phase Diagram Database of Japanese National Institute for Materials Science \cite{nims}}
\label{fig3}
\end{figure}

\section{The Cahn--Hillard equation}
\label{CH}

The Cahn--Hilliard equation plays a fundamental role in materials science, in description of the process of separation of the continuous medium into the regions with different concentrations.
It is naturally that this equation is also applicable in case of back processes. We start from the general form of Cahn--Hilliard equation  \cite{CH}, which has the following form:
\begin{gather}
\label{CHL1}
\frac{\partial c}{\partial t}=M_D{\nabla}^2\Big(\frac{\partial{\cal F}}{\partial c}\Big),
\end{gather}
where $c({\bf r},t)$ is the average concentration of solute, $M_D$ is transport coefficient (mobility) associated with diffusion coefficient, and
\begin{gather}
\label{CHF}
{\cal F}=f(c)+\frac12\varepsilon^2(\nabla c)^2,
\end{gather}
where $f(c)$ is the free energy density of the solution, and $\varepsilon^2(\nabla c)^2$ is the first non-vanishing term of the expansion of $f(c({\bf r}))$ in the Taylor series in ${\bf r}$, which describes the contribution of spatial correlation effects to free energy.
The cutting off the decomposition on the second-order term is equivalent to the assumption that the radius of intermolecular potentials action is much shorter than the characteristic lengths at which the concentration changes significantly. The molar volume is assumed to be independent of composition. $\varepsilon^2>0$ if the uniform state is stable at hight temperatures.

If $f(c)$ has a single minimum, then the equation (\ref{CHL1}) describes the usual diffusion with the characteristic dispersion relation \cite{SS} $\omega=-k^2(D+M_D\varepsilon^2k^2)$ between the frequency $\omega$ and the wave number $k$.  Here the gradient term contribution $(\nabla c)^2$ is assumed to be relatively small compared to the contribution of the free energy density. As a result one can express the diffusion coefficient is $D=M_D{\partial^2f(c)}/{\partial c^2}$.

It is important for us, that Cahn--Hilliard equation allows describe slow relaxation processes of spinodal decomposition in liquids. This process occurs when $f(c)$ is not linear and has two minima. Then, in the region between the minima, the diffusion coefficient $D<0$ and the dispersion relation $\omega=k^2(|D|-M_D\varepsilon^2k^2)$ indicate an important stabilizing role of the gradient term in the expression (\ref{CHF}). The negative diffusion coefficient in this expression corresponds to uphill diffusion, when the heterogeneities of the composition do not resolve, but are amplified by the reaction of the system. It leads to instability, and results in a characteristic ``worm-like'' structure of the solute distribution.  The relaxation processes in like systems are very slowly.

However,  the Gibbs potential of the Al--Y system, shown in Fig.\,\ref{fig3} has only single minimum. Therefore at the first blush the abnormally slow relaxation, be observed in Al--Y solution, has little in common with the spinodal decay. Nevertheless it should be note one feature of the eutectic Al--Y melt. It is the closely spaced vertical lines of stoichiometric compounds Al$_2$Y and Al$_3$Y. The physical processes taking place near peritectic, are quite complex and continue to be researched until the present time \cite{LFT,LUO}.  The presence of the compounds in initial solid can lead to existence in the melt of the local areas with a solute concentration exceeding the sample-average in the melt and corresponding to solid state in equilibrium. This should influence on diffusion and accordingly on the system relaxation dynamics which becomes nonlinear the same as in the case of spinodal decay. However, in our case the relaxation process should be in inverse way: from initial  non-homogenous state to full homogeneity.

In our case the effective Gibbs potential has more intricate form.
As it was note above in our initially non-homogenous binary melt the system's Gibbs potential is the combination of the potentials of liquid Al--Y solution and stoichiometric Al$_2$Y and Al$_3$Y compounds. Below, in order to construct an effective potential, we will use the corresponding Gibbs potentials.

\section{Relaxation in binary alloy with stoichiometry (phenomenological approach)}

Let us consider the model of a binary melt where $c$ is average concentration of impurity atoms in a basic liquid. We suppose that after melting this concentration is different in each point, and $c=\int c({\bf r}) dV$, where $c({\bf r})$ is the local concentration. This non-homogeneity is related with the presence in the initial sample of the stoichiometric phases inclusions, which are characterised by high impurity atoms concentration $c({\bf r})=c_c$ and corresponding  Gibbs energy $G_c$. We consider these inhomogeneities like remains of crystalline phase.
To describe the phase state of the solution we use the scalar field $\varphi$, such that for each unit of volume some share of the stoichiometric phase $\varphi$  corresponds, and $(1-\varphi)$ is the share of the liquid. If in the solid stoichiometric phase we assume $\varphi=1,$ then in the liquid phase this field is zero.
Therefore the impurity atoms concentration in the local volume is the sum of impurity concentration in liquid, $c_l({\bf r})$, and solid, $c_c$, phases:
\begin{gather}
\label{CL}
  c({\bf r})=(1-\varphi)c_l({\bf r})+\varphi c_c,
\end{gather}
and the Gibbs energy is the sum of the solid and liquid parts, $f(c)=G_c\varphi+G_l(c_l)(1-\varphi)$, where $G_l(c_l)$ is the concentration dependence of the Gibbs function of liquid Al--Y solution.

Unlike the known ideology of the phase field \cite{Elder}, the intermediate value of the field $0\le \varphi\le 1$ does not describe the interface between the phases, but corresponds to the volume mixture of phases in the spirit of the quasi-equilibrium theory of crystallization \cite{Flem}. Since we are not interested in the interface, we don't take into account the gradient contribution of the field $(\nabla\varphi)^2$, in contrast to the approach of the phase field, but consider the similar contribution to the liquid concentration of $c({\bf r})$. Neglecting the change in volume during phase transformations, we write down the system's molar free energy function in the following form:
\begin{gather*}
  {\cal F}=f_0+f(c)+\frac12\varepsilon^2(\nabla c)^2.
\end{gather*}

Binodal corresponds to the following condition:
\begin{gather*}
  \left.\dfr{\partial f}{\partial c_l}\right|_{c_l=c_b}=0, \left.\quad \nabla c\right|_{c_l=c_b}=0,
\end{gather*}
thus
\begin{gather*}
  \left.\dfr{\partial f}{\partial c_l}\right|_{c_l=c_b}=\left.(1-\varphi)\dfr{\partial G_l(c_l)}{\partial c_l}+\dfr{\partial \varphi}{\partial c_l}(G_c-G_l(c_l))\right|_{l_c=l_b}=0.
\end{gather*}
If we suppose that the full concentration is slow, then
\begin{gather*}
\dfr{\partial c}{\partial c_l}=(1-\varphi)+\dfr{\partial \varphi}{\partial c_l}(c_c-c_l)\approx 0,
\end{gather*}
and
\begin{gather*}
\dfr{\partial \varphi}{\partial c_l}\approx \dfr{1-\varphi}{c_l-c_c}.
\end{gather*}
In this case the binodal corresponds to the point $(c_b,\,T)$ on the phase diagram for which
\begin{gather*}-
(c_b-c_c)\left.\dfr{\partial G_l(c_l,\,T)}{\partial c_l}\right|_{c_l=c_b}\approx G_l(c_b,\,T)-G_c.
\end{gather*}
This expression allays to get usual phase diagram.

Note that the molar concentration of solute in the liquid $c_l({\bf r})$ is not a constant value and can vary both due to changes in the phase fraction and due to diffusion. The concentration $c({\bf r})$ per unit volume can be changed only by diffusion fluxes ${\bf J}_D$ in the liquid with a fraction $(1 - \varphi)$:
\begin{gather*}
\partial_t c({\bf r})=-(1-\varphi)\nabla\cdot{\bf J}_D,
\end{gather*}
the choice of which is due to the requirement to reduce the total Gibbs energy of the  system in the relaxation processes.
From the expression one can find the equation for solute in the liquid:
\begin{gather}
\label{eqx}
(1-\varphi)\partial_tc_l=-(c_c-c_l)\partial_t{\varphi}-(1-\varphi)\nabla{\bf J}_D.
\end{gather}
Here the diffusion flux in liquid is proportional to the gradient of molar chemical potential, $\mu ={\partial {\cal F}}/{\partial c_l}$:
\begin{gather*}
{\bf J}_D=-M_D\nabla \dfr{\partial {\cal F}}{\partial c_l},
\end{gather*}
where $M_D>0$ is the kinetic mobility coefficient.

\begin{figure}[h!]
   \centering
   \includegraphics[scale=0.65]{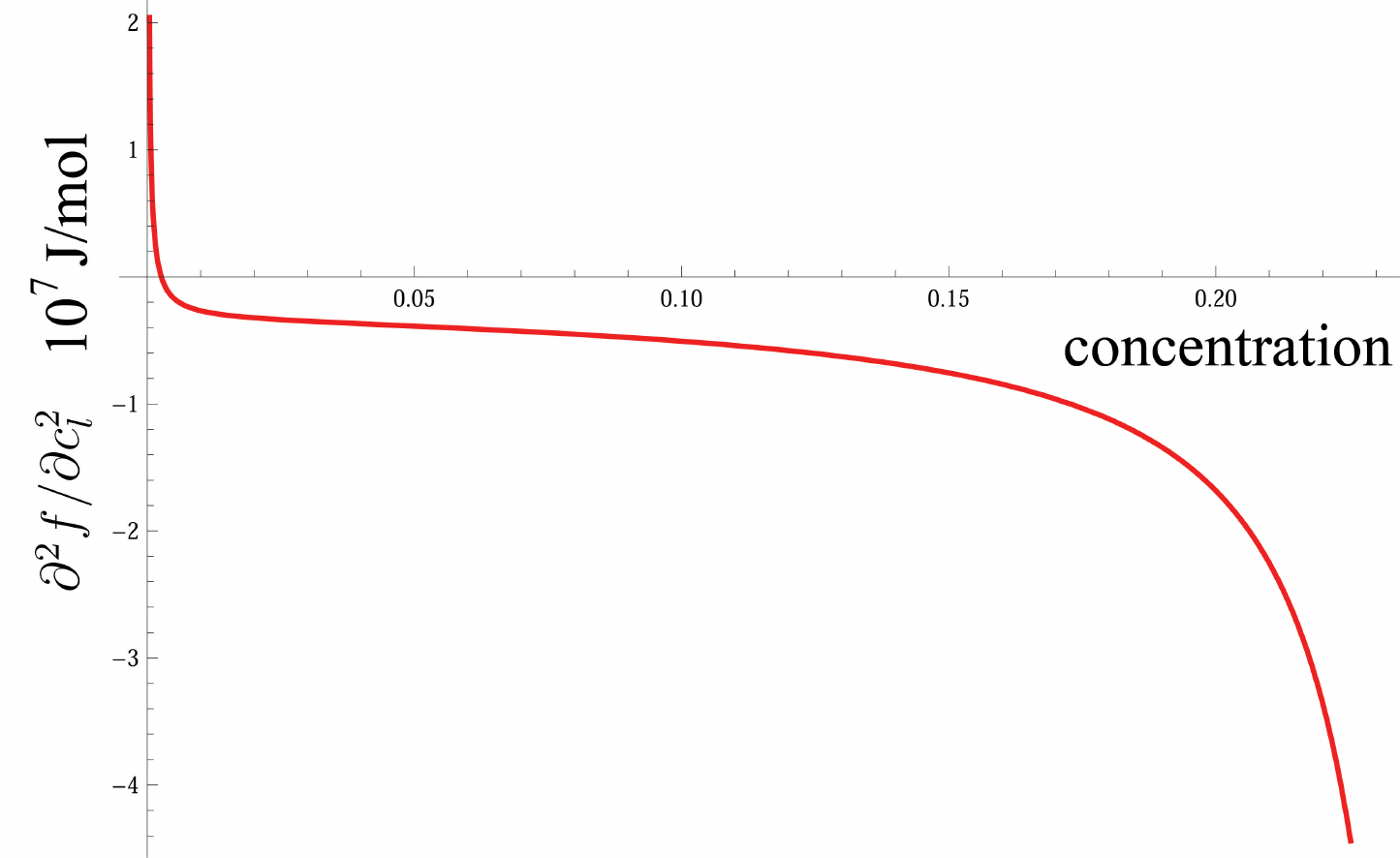}
   \caption{The calculated concentration dependence of the second derivative over $c_l$ of the Al--Y melt Gibbs energy at $T=1173$ K.}
   \label{F2}
\end{figure}

For simplification we suppose that the non-melted solid phase, $\varphi $, was evenly distributed equal in volume of initial solid sample, therefore $\nabla c({\bf r}) \approx (1-\varphi)\nabla c_l$, and the composition in every point is little different from average one, $\Delta c=c({\bf r})-c\ll 1$. Also it will be suppose that $\nabla \left({\partial^2f}/{\partial c_l^2}\right)=0$, that, off course, is strong approximation, but can be used for initiatory stage of relaxation. Then in expansion in series over $\Delta c({\bf r})$ one can confine oneself by the quadratic term, and approximated evolution equation has the following form:
\begin{gather}
\label{EQQ}
\partial_t c\approx M_D\left(\dfr{\partial^2f}{\partial c_l^2}\nabla^2c-\varepsilon^2\nabla^4c\right).
\end{gather}
In the above supposing that the full concentration is slowly changing, the second derivative of the free energy density has the following form:
\begin{multline*}
\dfr{\partial^2 f}{\partial c_l^2}\approx\dfr{\partial }{\partial c_l}\left[(1-\varphi)\dfr{\partial G_l(c_l)}{\partial c_l}+\dfr{\partial \varphi}{\partial c_l}(G_c-G_l(c_l))\right]=\\
(1-\varphi)\left[\dfr{\partial^2 G_l(c_l)}{\partial c_l^2}+\dfr{2}{(c_c-c_l)}\left[\dfr{G_c-G_l(c_l)}{c_c-c_l}+\dfr{\partial G_l(c_l)}{\partial c_l}\right]\right].
\end{multline*}
Hence now, using the Gibbs energy functions for the Al--Y liquid solution, and for stoichiometric Al$_2$Y and Al$_3$Y compounds from~\cite{nims} (Fig.\,\ref{fig3}), one can estimate the conditions of dynamical instability appearing. The calculated second derivative for considered Al--Y melt is shown in Fig.\,\ref{F2}, in which one can see that the region of concentrations, where the system relaxation dynamics can be instable, exists practically in all area of concentration.
Only the condition for manifestation of this instability is the initial melt non-homogeneity.

Usually, for homogeneous nucleation description the concept of the time-dependent structural factor $S({\bf k},t)$ is used \cite{SS}. The structural factor $S({\bf k},t)$ is proportional to square of the Fourier component,
\begin{gather*}
C({\bf k},\,t)=C({\bf k},0)e^{A({\bf k})t/2}=\dfr 1{(2\pi)^3}\int\limits_{\bf r}\left[c_l({\bf r},\,t)-c\right]e^{-i{\bf kr}}\mathrm{d}V,
\end{gather*}
of the order parameter, and usually increases as $S({\bf k},\,t)=\langle C^2({\bf k},\,t)\rangle\propto\exp(A({\bf k})t)$. The value $A({\bf k})$ is called ``the amplification rate''. Usually from experiments one finds dependence $A({\bf k}){\bf k}^{-2}$ \cite{SS}.
From the dispersion relation for the Cahn--Hilliard equation we can easily find that
\begin{gather*}
A({\bf k}){\bf k}^{-2} = -M_D\left(\dfr{\partial^2f}{\partial c^2}+\varepsilon^2{\bf k}^2\right).
\end{gather*}
gives a linear dependence by ${\bf k}^2$. Note that the experimental data on the the amplification rate are characterized by significant nonlinearity \cite{SS}. Nevertheless, it is believed that the Cahn--Hilliard equation correctly reflects the essence of what is happening in the spinodal decay: the convexity of the potential leads to the instability, which is extinguished by the fourth derivative.

In order to estimate the space scale, $\xi$, of initial inhomogeneities which is needed for start of non-linear system relaxation manifestation, we should solve the equation:
\begin{gather*}
\varepsilon^2\xi^{-2}=-\dfr{\partial^2f}{\partial c^2}.
\end{gather*}
The characteristic value of $\varepsilon^2$ is $10^{-7}$ Jm$^2$/mol, from Fig.\,\ref{F2} ${\partial^2f}/{\partial c^2}\sim 10^6$ J/mol. As a result the estimation gives $\xi\sim 10^{-6}$ m. It is enough to lead to the dynamic instability, since the initial heterogeneities scale is several orders of magnitude higher than obtained estimation (see Fig.\,\ref{LIG}).

\section{Solute dynamics in the liquid solution with stoichiometric compounds formation}

In order to study the solute redistribution processes between the liquid phase and stoichiometric compounds, resulting to the instability, we will limit ourselves to the convex potential model, for which it is possible to carry out up analytical calculations, and will analyze the dynamics of the system consisting both liquid and stoichiometric phases.
It is a reminder that in contrast with spinodal decomposition where the separation on regions with a different solute concentration but with one aggregate nature is discussed, here we consider the uphill diffusion in the processes of melting

Let us write down the free energy concentration of the liquid solution with stoichiometric compound inclusions in the following form:
\begin{gather}
\label{GS}
{\cal F}=f_0+\varphi G_c+(1-\varphi)G_l(c_l)+\frac12\varepsilon^2\big(\nabla c_l(1-\varphi)\big)^2,
\end{gather}
where in the last term we have taken into account the possibility of impurity diffusion only in liquid phase.

According to the nonequilibrium thermodynamics \cite{RR} the relaxation equations of the dynamical system, guaranteeing the decreasing of (\ref{GS}) functional, have the following form:
\begin{gather*}
\label{eq}
\displaystyle \dot{\varphi}=-M_{\varphi}\Big[G_c-G_l(c_l)-(c_c-c_l)\tilde{\mu}+c_l\varepsilon^2\nabla^2\big(c_l(1-\varphi)\big)\Big],\\[12pt]
\displaystyle {\bf J}_D=-M_D\nabla\Big[(1-\varphi)\tilde{\mu}\Big],
\end{gather*}
where $\tilde{\mu}=\mu-\varepsilon^2\nabla^2\big(c_l(1-\varphi)\big)$, and $M_{\varphi}>0$ is the kinetic coefficient of $\varphi$-field mobility which determines the rate of phase growth and is usually empirical.
Taking into account the conservation law (\ref{eqx}) one can write the equations determining the system dynamics in the following form:
\begin{multline}
\label{eq1}
\displaystyle \dot{\varphi}=M_{\varphi}\Big[G_l(c_l)-G_c+(c_c-c_l)\tilde{\mu}-c_c\varepsilon^2\nabla^2\big(c_l(1-\varphi)\big)\Big],\\[12pt]
\displaystyle (1-\varphi)\dot{c_l}=-(c_c-c_l)\dot{\varphi}+(1-\varphi)\nabla\Big(M_D\nabla\Big[(1-\varphi)\tilde{\mu}\Big]\Big).
\end{multline}

The obtained equations show that the phase transition velocity $\partial_t{\varphi}$ is determined by the difference between the grand potentials of the phases ($\Omega_l=G_l(c_l)-c_l{\mu}$, and $\Omega_c=G_c-c_c{\mu}$), that coincides with the accepted point of view about the driving forces of phase transformations (\cite{Pl,DLG}) in thermodynamic. The additional driving force of the phase transition is related with concentration correlations defined by $\nabla^2\left(c_l(1-\varphi)\right)$.  In areas where function $c_l(1-\varphi)$ is convex upwards, the stoichiometric phase grows, and where it is convex downward the stoichiometric phase decays.

\begin{figure}[h!]
   \centering
   \includegraphics[scale=0.63]{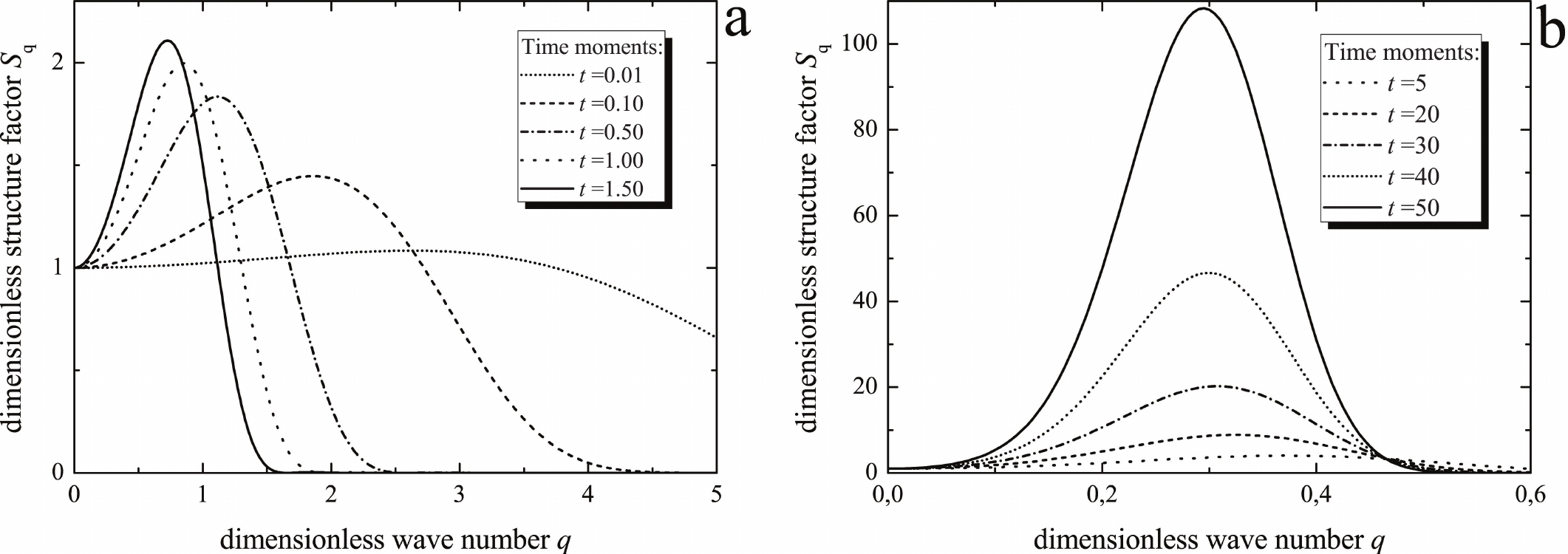}
   \caption{The model structure factor change in time a) initial stage, b) final stage~\cite{LV}}
   \label{R1}
\end{figure}

The equation system (\ref{eq1}) can be linearized and solved numerically (\ref{eq1})~\cite{LV} using the Gibbs potentials of Al--Y system from Computational Phase Diagram Database of Japanese National Institute for Materials Science (NIMS) \cite{nims}. The calculated time-dependent structural factor of the system , $S({\bf k},\,t)=N_q|c_l^2({\bf k},\,t)|$ ($N_q$ is the normalization factor), has the form presented in Fig.\,\ref{R1}.

The initial state of the considered system in our calculations was inhomogeneous state with uniform distribution of the inhomogeneities sizes. The presented results show that within relaxation in the system the coalescence process is observed, where the large-scale inhomogeneities grow at the expense vanish of small fluctuations. As we have noted above this implies uphill diffusion, or a negative diffusion coefficient. The grow of the fluctuations size is limited by the value defined by ratio of volume energy of inhomogeneity to interface energy, that clear viewed in our results.

We have to note, that from our calculations it follows that at final stage the number of these inhomogeneities increase without limit, that actually is not correct and contradicts to experiment, where inhomogeneity vanishes in equilibrium state. However, this contradiction is natural since in our model we do not take into account the impurity atom number limit restricting the inhomogeneities growth, as well as the thermal fluctuations destroying them.
Therefore, we are aware it does not accurately describe the relaxation process, and can be used only for qualitative estimation of the system dynamics on initial stage of relaxation. In over hand, the coalescence is observed in our model also at the initial stage of relaxation when the approximation is still applicable. Thus, we can suppose that the described above instability significantly does influence on the relaxation dynamics and is the cause of extremely large relaxation time.

\section{Estimation of relaxation dynamics in fluctuation region}

As we noted above, at large observation time the thermal fluctuations play important role, leading the system to a thermodynamically equilibrium state. This relaxation stage can be analytically described using the fluctuation theory of phase transitions and non-equilibrium dynamics methods.

In our case of Al--Y melt the instability leads to extremely large relaxation time. Indeed, the second and fourth terms of the Gibbs function expansion in Taylor series over $\Delta c({\bf r})$ are close in value module and opposite in sign:
\begin{gather*}
\dfr{\partial^4 f(c)}{\partial c^4}\sim 10^{7}\,\mbox{J/mol}, \quad \dfr{\partial^2 f(c)}{\partial c^2}\sim -10^{6}\,\mbox{J/mol}.
\end{gather*}
Therefore, the effective amplification rate at large time becomes close to zero:
\begin{gather*}
A\approx -\xi^{-2}M_D\left(\dfr{\partial^4 f(c)}{\partial c^4}\langle (\Delta c)^2\rangle+\dfr{\partial^2 f(c)}{\partial c^2}\right)\approx 0.
\end{gather*}

The instability can explain not only slow relaxation of the eutectic melts, but also its non-monotonic character. In order to illustrate our statement we use results of the fluctuation dynamic theory of phase transition applied to the ``toy model'' of a system undergoing an weak first order phase transition \cite{VMI}. For this we firstly write the time-dependent correlation function corresponding to equation (\ref{EQQ}):.
\begin{gather*}
  \langle c^2\rangle_{{\bf k},\,t}= \dfr{\exp\left[-M_D{\bf k}^2\left(\varepsilon^2{\bf k}^2+m^2\right)|t|\right]}{{\bf k}^2+m^2},
\end{gather*}
where
\begin{gather*}
m^2=\left(\dfr{\partial^4 f(c)}{\partial c^4}\langle (\Delta c)^2\rangle+\dfr{\partial^2 f(c)}{\partial c^2}\right).
\end{gather*}

\begin{figure}[h!]
   \centering
   \includegraphics[scale=0.8]{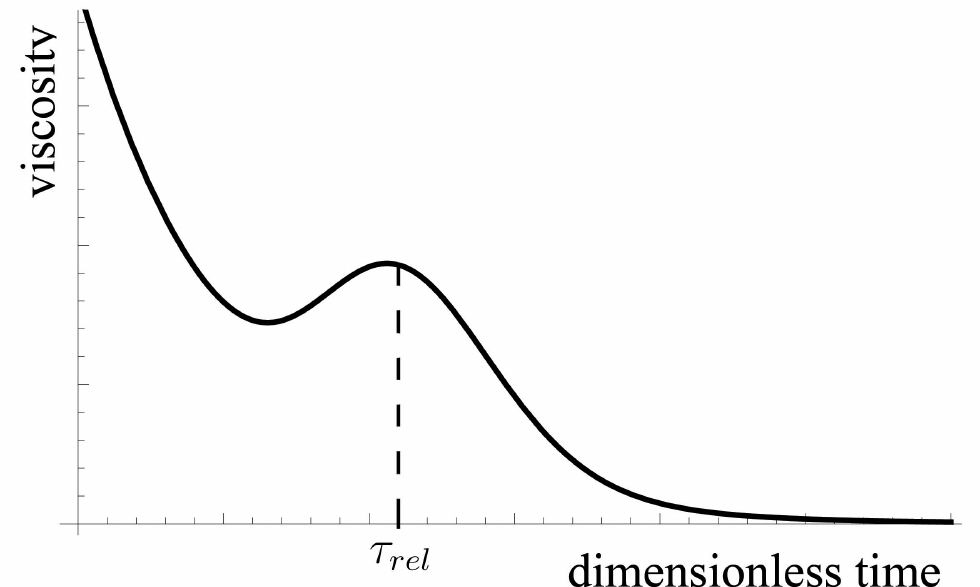}
   \caption{Qualitative form of viscosity time dependence corresponding to (\ref{NMR})}
   \label{R10}
\end{figure}

In the fluctuation region near the binodal, where $m^2\gtrsim 0$, the thermal fluctuations sensibly influence to relaxation dynamics, and expression for time-dependence viscosity of the melt can be presented in following form~\cite{VMI}:
\begin{gather}
\label{NMR}
  \eta (t)\propto\eta_0+\eta^*\int\limits_{V_{\bf k}} G_R({\bf k},\,t)\int\limits_t^{\infty}G_R({\bf k},\,\tau)\mathrm{d}\tau\mathrm{d}{\bf k},
\end{gather}
where
\begin{multline*}
  G_R({\bf k},\,t)=\exp\left[-t/\tau_{rel}+C\erf(-\sqrt{t/\tau_{rel}})-\right.\\ \left.
  2C\sqrt{2t/\pi\tau_{rel}}e^{-t/\tau_{rel}}-M_D{\bf k}^2(\varepsilon^2{\bf k}^2+m^2)t\right],
\end{multline*}
and $\tau_{rel}\sim \xi^2(M_Dm^2)^{-1}$ is the relaxation time.
The qualitative form of this function agree with experimentally observed non-monotonic time-dependence  of viscosity (compare Fig.\,\ref{R10} with Fig.\,\ref{fig1}).

\section{Conclusions}

In conclusion we summarise our statements. The main idea of our work is very simple: We think, and had tried to show here, that slow and non-monotonic relaxation processes in some melts with eutectic composition are possible because of initial melt strong heterogeneity combined with the Gibbs function nonlinearity induced by the proximity of stoichiometric compound on the phase diagram.

In order to study the solute redistribution processes between the liquid phase and remains of the stoichiometric compounds, resulting to the instability, we limited ourselves to the convex potential model, for which it is possible to carry out up analytical calculations, and analyzed the dynamics of the system consisting both liquid and stoichiometric phases.

For performance one can conclude that the nature of slow relaxation processes in eutectic melts after melting is similar with the nature of spinodal decomposition. In both cases the slowing reason is the system thermodynamic instability. The difference is that in the spinodal decomposition because of the thermodynamic instability the development of inhomogeneous heterogeneous structures occurs, and in the slow relaxation processes in presence of the thermodynamic instability an initially heterogeneous structure slowly relaxes to homogeneous state.

Of course, in spite of presence of the thermodynamic instability and positive value of the amplification rate the inhomogeneous heterogeneous structures will disperse with time, when the system will reach equilibrium state. In order to describe in details this relaxation dynamics in terms of Cahn--Hilliard equation one should take into account the amplification is extinguished by the fourth term of the Gibbs function expansion in Taylor series and by the account of thermal fluctuations. We shown this allows qualitatively describe non-monotonic relaxation observed in some Al-based eutectic melts.

\section*{Acknowledgments}

The work was supported by Russian Foundation for Basic Research, Grants 18-02-00643 (MV) and 18-42-180002 (VL). Part of the work was carried out within the framework of the state assignment of the Ministry of Education and Science of Russia (No.AAAA-A17-117022250039-4)

%%%%%%%%%%%%%%%%%%%%%%%%%%%%%%%%%%%%%%%%%%%%%%%%%%%%%%%%%%%%%%%

\end{document}